\documentclass{article}

\usepackage{PRIMEarxiv}

\usepackage[utf8]{inputenc} 
\usepackage[T1]{fontenc}    
\usepackage{hyperref}       
\usepackage{url}            
\usepackage{booktabs}       
\usepackage{amsfonts}       
\usepackage{nicefrac}       
\usepackage{microtype}      
\usepackage{lipsum}
\usepackage{fancyhdr}       
\usepackage{graphicx}       
\graphicspath{{media/}}     

\usepackage{cite}
\usepackage{amsmath,amssymb,amsfonts}
\usepackage{algorithmic}
\usepackage{graphicx}
\usepackage{textcomp}

\usepackage{tikz, graphicx, mathtools, amsfonts, mathrsfs, amsmath, xfrac, amssymb, dsfont, enumitem,bbold, multirow}

\usepackage[ruled,vlined]{algorithm2e}
\SetKwInOut{Parameters}{Parameters}
\SetKwInOut{Initialization}{Initialization}

\DeclareMathOperator*{\argmax}{arg\,max}
\DeclareMathOperator*{\argmin}{arg\,min}

\newcommand{\naturals}{\mathbb{N}}
\newcommand{\reals}{\mathbb{R}}

\newcommand{\red}[1]{\textcolor{black}{#1}}
\newcommand{\blue}[1]{\textcolor{black}{#1}}

\newcommand{\norm}[2]{\left\lVert#1\right\rVert_{#2}}

\newcommand{\reach}[3]{\mathcal{R}^{#1}\left({#2};{#3}\right)}

\newtheorem{remark}{Remark}

\newtheorem{proof}{Proof}
\newtheorem{theorem}{Theorem}

\title{Approximating Reachable Sets for Neural Network based Models in Real-Time via Optimal Control
\thanks{© 2022 IEEE. Personal use of this material is permitted. Permission from IEEE must be obtained for all other uses, in any current or future media, including reprinting/republishing this material for advertising or promotional purposes, creating new collective works, for resale or redistribution to servers or lists, or reuse of any copyrighted component of this work in other works.\\
Omanshu Thapliyal is a PhD candidate with the School of Aeronautics \& Astronautics Engineering at Purdue University, West Lafayette, IN 47906. \\
Inseok Hwang is a professor at the School of Aeronautics \& Astronautics Engineering at Purdue University, West Lafayette, IN 47906.}}

\author{
  Omanshu Thapliyal\\
  Purdue University \\
  West Lafayette, IN USA\\
  \texttt{omanshu@purdue.edu} \\
   \And
  Inseok Hwang \\
  Purdue University \\
  West Lafayette, IN USA\\
  \texttt{ihwang@purdue.edu} \\
}

\begin{document}
\maketitle

\begin{abstract}
In this paper, \blue{we present a data-driven framework for real-time estimation of reachable sets for control systems where the plant is modeled using neural networks (NNs).}
We utilize a running example of a quadrotor model that is learned using trajectory data via NNs.
The NN learned offline, can be excited online to obtain linear approximations for reachability analysis.
We use a dynamic mode decomposition based approach to obtain linear liftings of the NN model.
\blue{The linear models thus obtained can utilize optimal control theory to obtain polytopic approximations to the reachable sets in real-time.
The polytopic approximations can be tuned to arbitrary degrees of accuracy.}
The proposed framework can be extended to other nonlinear models that utilize NNs to estimate plant dynamics.
We demonstrate the effectiveness of the proposed framework using an illustrative simulation of quadrotor dynamics.
\end{abstract}


\section{Introduction}
\label{sec:introduction}
As the systems of interests of control engineers get more complex, and data get inexpensive to obtain in large quantities, machine learning finds increasing applications in control systems.
Further, obtaining simulated data for multiple trajectories of the system of interest is often easier than designing physical control.
For instance, neural networks (NNs) often find applications to model plants, actuators, controller logic, and even for modeling the human operator's intent and logic.
To this end, data-driven approaches to discover underlying physical models for dynamical systems are very useful.

\blue{Besides, set-based properties of safety, reachability, and controllability provide strong analytical bases to quantify system performance (especially under uncertainties).
Of these, estimating reachability property is closely tied with other properties such as viability, controllability and safety \cite{lygeros2004reachability}.}
Additionally, reachability can be utilized for optimal control synthesis and high level decision making \cite{ahn2020reachability}.
Reachable sets can be computed analytically by solving Hamilton Jacobi (HJ) partial differential equations (PDEs).
\blue{Such HJ based solutions of reachable sets are time consuming, and suffer from the `curse of dimensionality', making real-time applications difficult.}
To alleviate this, various numerical approximation techniques have been proposed to compute approximate reachable sets without explicit solutions to the associated HJ PDEs.
These techniques employ polytopic approximations \cite{hwang2005polytopic}, Pontryagin's optimal control \cite{varaiya2000reach}, and numerical differential equation solvers \cite{dit2018reachability}, 
to name a few. 
\blue{Based on these, a number of reachable set computation tools are available that utilize numerical techniques for approximate reachable sets and tubes (such as \cite{1281781, althoff2018implementation, immler2018arch}), both forward and backwards in time.}

\blue{On the other hand, the ability to design control signals, and comment upon system properties under operational noise, parameter uncertainties, and system nonlinearities, is of a lot of importance.
As a result, NNs are applied in the entire {control system design process} of system identification \cite{romeres2019derivative}, output tracking \cite{devasia2017iterative}, control synthesis \cite{vaupel2020accelerating}, state estimation \cite{cheng2019hidden}, and devising supervisory control logic \cite{thapliyal2021}.
Recently, NNs have been applied to learn nonlinear dynamical models of varying complexity.
A robotic arm's inverse dynamics are inferred in \cite{romeres2019derivative} using iterative learning on real data from an iCub robotic arm.
In \cite{thapliyal2021}, a learning based scheme is used to infer supervisory control logic for cybersecurity analysis of supervisory control systems.
The authors synthesized control signals to control a quadrotor by learning its dynamics in \cite{bansal2016learning}.
More generally, utilizing machine learning techniques for system identification has been noted to be particularly useful in numerous recent system modeling and identification texts such as \cite{chiuso2019system}, \cite{pillonetto2014kernel}, and \cite{garg2017system}.}
\blue{However, \emph{despite the widespread usage of NNs} in solving dynamics and control problems, \emph{the absence of reachable set computation/approximation tools for NN based models prohibits a reliable application} of NN under uncertainties and operating conditions that demand safety guarantees.}
Since the machine learning models use data from an unknown dynamical system, numerical approaches to compute approximate reachable sets can be extended to the learned models themselves.
\blue{To this end, Koopman Operator theory has been used to learn NNs as it provides a method to find a computationally scalable, equivalent linear lifted model \cite{folkestad2021koopman}.
Conversely, learning-based methods are also used to learn the Koopman operator itself, for control synthesis \cite{han2020deep}. 
Nevertheless, linear control methods can be utilized on such linear lifted models to estimate reachable sets as polytopes, and an optimal control problem can be formulated to propagate these polytopes over time.}
The polytopic reachable set approximation can be made arbitrarily accurate \cite{varaiya2000reach}, therefore, the reliability of the proposed method is conditioned on the accuracy of: a) the NN being able to approximate unknown plant models, and b) the Koopman operator {being} able to lift the NN model to a higher dimensional manifold and approximate it as a linear system.

\subsection{Related Works}\label{sec:intro-relatedworks}
The general problem of computing output bounds of a NN can be readily related to reachable set computation for learned dynamical models.
Output bound computation for NN based controllers using specific activation functions, has been achieved by solving mixed-integer linear programs (MILPs) \cite{dutta2018output} and relaxed linear programs (LPs) \cite{xiang2018reachability}.
This makes the amenability towards real-time applications particularly low.
Such methods exploit the properties of individual perceptrons in a NN, connected via rectified linear unit (ReLu) activations to obtain output bounds.
{The} authors in \cite{huang2019reachnn} employ Bernstein polynomials to obtain Taylor approximations of NN based control systems to obtain reachability flow pipes.
In \cite{xiang2018reachable}, exact reachable sets are computed for a control system employing ReLu activation functions based NNs, with specific switched linear dynamics.
The above methods utilize explicit system dynamics, or specific activation functions to obtain reachable sets for NN models.
Most NNs employed to learn system dynamics can be arbitrarily nonlinear. 
\blue{On the other hand, reachable set computation/approximation using HJ methods or polytopes is extant in controls literature \cite{lygeros2004reachability,hwang2005polytopic,varaiya2000reach}.
Being exact model-based methods, they rely on the complete knowledge of the dynamical modes of the system, an assumption no longer true for NN based system modeling.}
To the best of our knowledge, there do not exist methods that extend optimal control theory based on local linear system approximations of the given NN to obtain approximate reachable sets \blue{for the} NN models.

\subsection{Contributions}
The main contributions of the proposed method are listed as follows.
1) We utilize a dynamic mode decomposition based framework to obtain approximate linear models for the given learned nonlinear dynamics.
This allows us to use optimal control theory to obtain polytopic approximations to the reachable sets for the approximately equivalent linear system.
2) The proposed framework is numerically efficient, and much more amenable to real-time applications than solving MILPs or relaxed LPs at each time step.
\blue{We demonstrate this using a realistic and detailed example of real-time reachable set approximation for a quadrotor model --- a widely studied system for identification and control using NNs.
3) Finally, the framework can employ `plug-and-play' reachability modules from other reachability assessment tools for the approximately equivalent linear systems.
That is, the introduced polytopic reachable set approximation methods can be replaced by other reachability modules, as shown in Fig.\ref{fig:schematic}.}

The rest of this paper is organized as follows.
In Section 2, we formulate the reachability problem for a learned model.
Section 3 contains the main framework to estimate reachable sets for nonlinear models learned by a NN.
The approximate reachable set computation is posed as an optimal control problem to obtain polytopic reachable sets.
In Section 4, we implement the reachability estimation framework on an illustrative quadrotor example.
We {first consider} a nominal quadrotor reachability case, and {then} consider a separate scenario of estimating reachable sets when two of the rotors have failed. 
Finally, {Section 5} presents our concluding remarks.

\textit{Notations: }
For two vectors $u$ and $v$, their inner product is denoted by $\langle u, v\rangle$.
For a matrix $A$, we {denote} its transpose by $A^T$ and its Frobenius norm by $\norm{A}{F}$.
For two sets $A$ and $B$, we denote their Minkowski sum as $A\oplus B \triangleq \{a+b\, \lvert\, a\in A,b\in B\}$. 
For a finite set $A$, if a random variable $x$ is distributed uniformly in the set $A$, we write $x\sim \mathcal U_{A}$.

\section{Problem Formulation}
\blue{Consider {the} nonlinear dynamical system given as follows:
\begin{equation}\label{eq:nl-sys}
\dot{x}(t)=f\left(x(t),u(t)\right),\, \mathrm{ and }\;x(0)\in\mathcal X_0,u(t)\in\Omega, \;t\geq 0
\end{equation}
where $x\in\mathcal X\subseteq \reals^{n_x}$ is the state, $x_0$ is the initial state in a known initial set $\mathcal X_0$, and the control input $u\in\reals^{n_u}$ resides in the set $\Omega$ at all times $t$.
Given some input-state data in the form of $X_k\triangleq \{x_0,\cdots,x_N\}$, $U_k\triangleq \{u_0,\cdots,u_N\}$ over multiple trajectories $k=0,\cdots,n_T$, the unknown dynamical map $f$ is learned using the trajectory data.
The trajectory data is sampled from $\dot{x}=f(x,u)$ in (\ref{eq:nl-sys}) at some sampling rate $\Delta t$, such that $\left(x_i,u_i\right)\triangleq\left(x(i\Delta t),u(i\Delta t)\right)$ for $i=1,\cdots,N$.}

\blue{A data-driven method, such as a neural network (NN), is employed to estimate the unknown dynamics $f$ from the time series data trajectories as:}
\begin{equation}\label{eq:nn-sys}
\dot{\tilde{x}}(t)=\tilde{f}_{\Theta}\left(\tilde{x}(t),u(t)\right)
\end{equation}
Here $\tilde{x}(t)$ is the state obtained by the NN from the data $X_k,U_k$, and it approximates the true state $x(t)$ as long as $\tilde{f}_\Theta\approx f$.
The approximate dynamics $\tilde{f}_\Theta$ is parameterized by $\Theta$ which contains the parameters of the learning method employed (i.e., NN weights and biases).
Without loss of generality, we assume the initial state $\tilde{x}(0)\in\mathcal X_0$ and control input $u\in\Omega$.
The reachability problem for the NN model is then to find:
\begin{equation}\label{nn-reach}
\reach{\tilde{f}}{\tau}{\mathcal X_0} \triangleq \left\{\tilde{x}(\tau) \, \lvert\, \dot{\tilde{x}}=\tilde{f}_{\Theta}\left(\tilde{x},u\right),\tilde{x}(0)\in\mathcal X_0,u\in\Omega; \Theta\right\}
\end{equation}
at some time $\tau$, given initial conditions $\mathcal X_0$ and admissible control set $\Omega$.
\blue{Note that the problem to compute $\reach{\tilde{f}}{\tau}{\mathcal X_0}$ is nontrivial due to the arbitrary, unstructured nonlinearities present in the NN modeling (parameterized by $\Theta$).}
\begin{remark}[NNs as Universal Approximator \cite{sonoda2017neural}]
Given time series trajectory data $X_k,U_k$, a causal NN with parameters $\Theta$ can be employed to approximate $f$ appropriately.
Obviously, $\tilde{f}_\Theta \to f$ as the number of available trajectories $n_T\to\infty$, therefore, $\tilde{x}(\tau)\to x(\tau)$ for $0\leq \tau \leq N\Delta t$.
\end{remark}

So, if we can find the set $\reach{\tilde{f}}{\tau}{\mathcal X_0}$, it gives us a good approximation of the reachable set for the nonlinear system in (\ref{eq:nl-sys}).
Since we do not concern ourselves with devising a new learning scheme, we shall restrict our discussion to $\reach{\tilde{f}}{\tau}{\mathcal X_0}$.
\blue{Moving forward, \emph{we assume the dynamics of the system to be given by $\tilde{f}_{\Theta}$, as the NN can be trained accurately due to Remark 1.}}

\section{Reachability Framework for Neural Network Models}
In this section, we will assume the nonlinear dynamical system has been learned using available NN techniques and focus on approximating reachable sets of NN dynamics $\tilde{f}_\Theta$ using a relatively computationally cheap method.
To this end, we revise a formulation of dynamic mode decomposition (DMD) that allows for control inputs \cite{proctor2016dynamic}.
This allows us to build finite-dimensional approximation to the infinite dimensional Koopman operator, to get equivalent linear time-varying system models. 

\begin{figure}[!t]
    \centering
    \includegraphics[width=0.85\columnwidth]{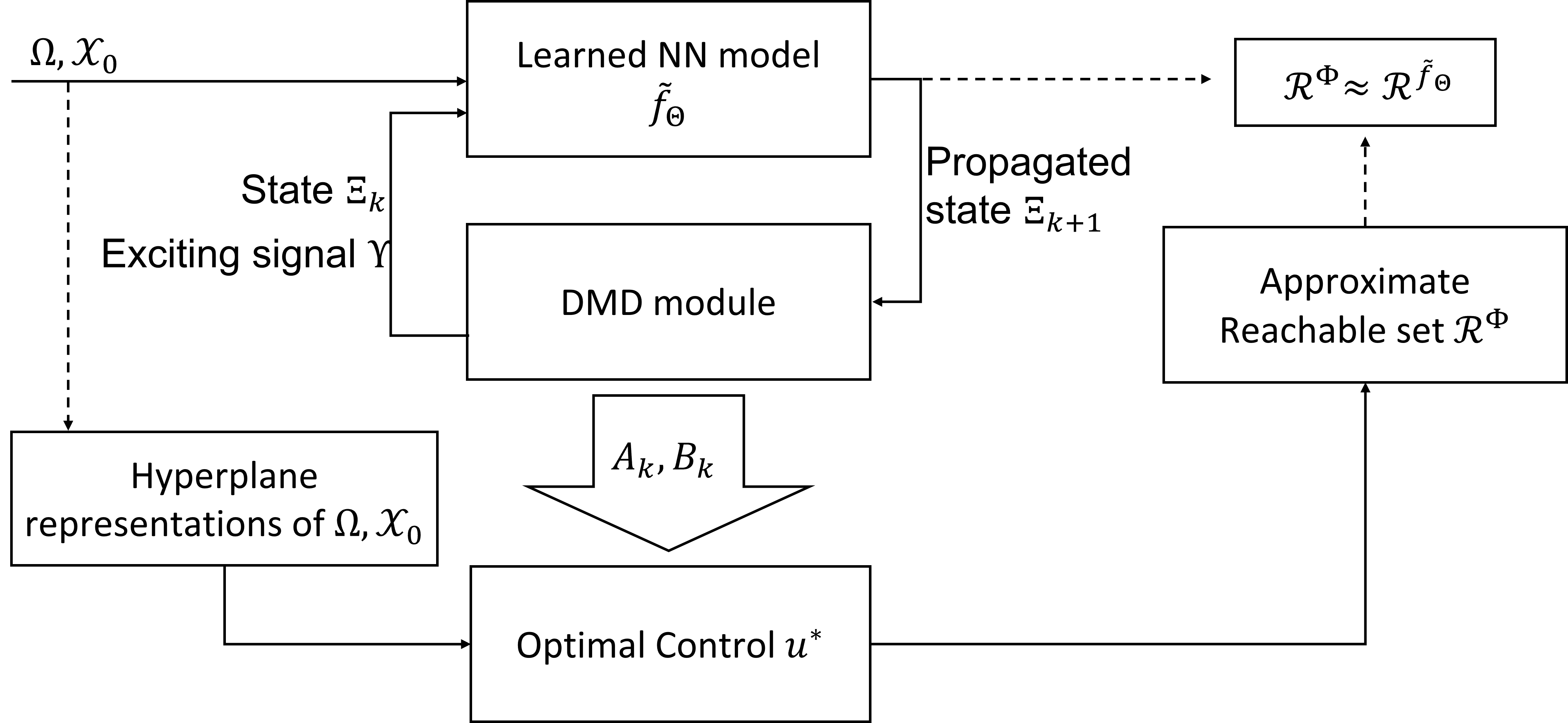}
    \caption{A schematic of the proposed data-driven framework for approximate reachable set computation}
    \label{fig:schematic}
\end{figure}

\subsection{Dynamic Mode Decomposition with Control}
Let us look at a data-driven method for approximating the Koopman Operator, called dynamic mode decomposition.
Nominal forms of DMD involve trajectory data (called `snapshots') consisting of state evolutions over time $x_k$.
The trajectory data get mapped under a linear operator as $x_{k+1}\approx A x_k$.
DMD with control (DMDc) was proposed to include input-state relations to such trajectory evolutions in \cite{proctor2016dynamic}.
Given an input-state data point $x_k,u_k$, DMDc attempts to find the pair of operators $A,B$ such that $x_{k+1}\approx A x_k + Bu_k$ for data points on state $x_k\in\reals^{n_x}$ and input $u_k\in\reals^{n_u}$.
The data matrices at {time step $k$} are temporal snapshots of the trajectory, of width $w\in\naturals$, given by:
\begin{equation}
\Xi_{k,w} \triangleq \begin{bmatrix}
\vert & & \vert\\
x_k & \cdots & x_{k+w}\\
\vert & & \vert
\end{bmatrix}, 
\Upsilon_{k,w} \triangleq \begin{bmatrix}
\vert &  & \vert\\
u_k  & \cdots & u_{k+w}\\
\vert &  & \vert
\end{bmatrix} \end{equation}
The snapshot with data points propagated one-step in time, can then be represented as:
\begin{equation}\label{eq:dmdc}
\Xi_{k+1,w} = \Gamma_{k,w}\,\begin{bmatrix}
 \Xi_{k,w} \\ \Upsilon_{k,w}
\end{bmatrix}, \text{ where }\Gamma_{k,w}\triangleq \begin{bmatrix}
A & B
\end{bmatrix}
\end{equation}
Note that the mapping $\Gamma$ varies over time, and is parameterized by the snapshot width $w$.
The DMDc solution to (\ref{eq:dmdc}) can be viewed as a least-square regression problem to find a $\Gamma\in\reals^{n_x\times(n_x+n_u)}$ such that:
\begin{equation}\label{eq:ls-dmd}
\begin{split}
\Gamma_{k,w} &= \argmin{\norm{\Xi_{k+1,w} -  \Gamma_{k,w}\,\begin{bmatrix}
 \Xi_{k,w} \\ \Upsilon_{k,w}
\end{bmatrix} }{F}}, \text{ or}\\
\Gamma_{k,w} &=  \begin{bmatrix}
A & B
\end{bmatrix} = \Xi_{k+1,w} \begin{bmatrix}
 \Xi_{k,w} \\ \Upsilon_{k,w}
\end{bmatrix}^{\dagger}
\end{split}    
\end{equation}
where $[\cdot]^\dagger$ denotes the pseudo-inverse.
Extracting columns from the least-square solution in (\ref{eq:ls-dmd}), we get a linear time-varying system such that $x_{k+1}\approx A_k x_k + B_k u_k$.

A numerically efficient way to compute the least-square solution utilizes singular-value decompositions of data snapshot matrices \cite{proctor2016dynamic}.
Over time, matrices $A_k,B_k$ can be estimated in real-time with a time-moving window of width $w$.
In our case, the moving window obtains snapshot data from the input-state relation learned by the NN.
Note that the state snapshot data can be easily obtained by exciting the learned model $\tilde{f}_{\Theta}$ by sampling an arbitrary control input from $\Omega$ that forms $\Gamma_{k,w}$, and noting NN output into the propagated state data snapshot $\Xi_{k+1,w}$.
This gives an independent framework that uses excitations of the NN model to obtain approximate linear models, depicted in Fig.~\ref{fig:schematic}.

\subsection{Approximate Reachable Set Computation using DMDc Model}
Now that we have linear approximations of the form $x_{k+1}=A_kx_k + B_k u_k$, we will focus on obtaining reachable sets for the {linear time-varying} system $(A(t),B(t))$. 
Here, the system matrices $A(t)$ and $B(t)$ satisfy $\exp{\{A(t)\Delta t\}}=A_k$ and $\int^{\Delta t}_0 {\exp{\{A(s)s\}}B(s)ds}=B_k$ for $t\in[k\Delta t, (k+1)\Delta t)$.
That is, $A_k$ and $B_k$ get updated at time step $k+1$ upon receiving new snapshot data via (\ref{eq:dmdc}), hence, $A(t)$ and $B(t)$ get updated at time $(k+1)\Delta t$.
Let $\Phi(t,0)$ be the state-transition function associated with the linear system $(A(t),B(t))$.
That is, $\dot{\Phi}(t,0)=-A(t) \Phi(t)$, and $\Phi(0,0) = I$.
Let $\reach{\Phi}{\tau}{\mathcal X_0,\Omega}$ be the reachable set of the system $(A(t),B(t))$ at time $\tau$.
If the DMDc approximation is accurate, it should suffice to concern ourselves with approximating $\reach{\Phi}{\tau}{\mathcal X_0,\Omega}$.

Without loss of generality, we assume that the admissible control set and the initial state set are polytopes as:
\begin{equation}\label{eq:polytope-def}
\begin{split}
\mathcal X_0 &= \bigcap_{i=1}^{n_{1}} {\{v\in\reals^{n_x}\, \lvert\,\langle c_i(0), v\rangle \leq \gamma_i(0)\}}\\
\Omega &= \bigcap_{i=1}^{n_{2}} {\{u\in\reals^{n_u}\, \lvert\,\langle d_i, u\rangle \leq \varepsilon_i\}}\\
\end{split}
\end{equation}
\red{defined for arbitrary vectors $v$.
Additionally, $c_i$ and $d_i$ are normal vectors parameterizing the hyperplanes defining each face of the polytopes in (\ref{eq:polytope-def}).
Let the hyperplanes $H_i\equiv \langle c_i(0), v\rangle = \gamma_i(0)$ touch the set $\mathcal X_0$ at points $x^*_i(0)$ for $i=1,\cdots,n_1$.}
The above polytopic assumption is not limiting, as arbitrarily convex, compact sets $\mathcal X_0$ and $\Omega$ can be bounded by tight polytopes as (\ref{eq:polytope-def}).
Clearly, the following hold true:
\begin{equation}\label{eq:plane-contact}
\begin{split}
x^*_i(0) &=\argmax_{v\in \mathcal X_0}{\{\langle c_i(0),v\rangle\}}\text{, and} \\
\gamma_i(0) &= \max_{v\in\mathcal X_0} {\{\langle c_i(0),v\rangle \}}
\end{split}    
\end{equation}
A polytopic reachable set approximation looks at only the points of contact $x^*_i(0)$ of reachable sets of the linear system (see \cite{hwang2005polytopic,varaiya2000reach}).
Similarly, let $x^*_i(\tau)$ be the point of contact of $\reach{\Phi}{\tau}{\mathcal X_0,\Omega}$ at the hyperplane $H_i(\tau)$ defined as:
\begin{equation}
H^*_i(\tau) = \{x\,\lvert\,\langle c_i(\tau),x^*_i(\tau)\rangle = \gamma_i(\tau)\}
\end{equation}
for $i=1,\cdots,n_1$, at time $\tau>0$.
Such an argument can be made for compact and convex sets $\mathcal X_0$ and $\Omega$.
From \cite{varaiya2000reach}, this ensures that the reachable set $\mathcal R^{\Phi}$ remains compact and convex at all times.
This allows us to find hyperplanes $H^*_i(\tau)$ to support the reachable set.
Using an argument similar to (\ref{eq:plane-contact}), the point of contact between $H^*_i(\tau)$ and $\mathcal R^\Phi$ at an arbitrary time $\tau$ can be defined as:
\begin{equation}\label{eq:x_reach-contact}
\begin{split}
x^*_i(0) &=\argmax_{v\in  \reach{\Phi}{\tau}{\mathcal X_0,\Omega} }{\langle c_i(\tau),v\rangle}\\ 
&= \argmax \Big\{\langle c_i(\tau),v\rangle\,\text{ s.t. } v(t)= \Phi(t) x(0) + \int^t_0 {\Phi(t,s)B(s)u(s)ds}, u(t)\in\Omega,t\leq \tau\Big\} 
\end{split}
\end{equation}
Also, from (\ref{eq:plane-contact}), the distance between $H^*_i(\tau)$ and $\mathcal R^\Phi$ can be expressed as:
\begin{equation}\label{eq:gamma_reach-contact}
\begin{split}
\gamma_i(\tau) &= \max_{v\in \reach{\Phi}{\tau}{\mathcal X_0,\Omega} } {\langle c_i(\tau),v\rangle}\\
&= \max \Big\{\langle c_i(\tau),v\rangle\,\text{ s.t. } v(t)= \Phi(t) x(0) +  \int^t_0 {\Phi(t,s)B(s)u(s)ds}, u(t)\in\Omega,t\leq \tau\Big\} 
\end{split}
\end{equation}
For a given set of initial points of contact $x^*_i(0)$, equations (\ref{eq:x_reach-contact}) and (\ref{eq:gamma_reach-contact}) depend only on the choice of $u\in\Omega$, thereby forming an optimal control problem.

\begin{theorem}
Let the optimal control $u^*_i(\tau)$ be the solution to $\argmax_{u(\tau)}{\langle c_i(\tau), B(\tau) u(\tau)\rangle}$.
Then, for $\dot{c_i} = -A(\tau)^T c_i(\tau)$ with the initial condition $\gamma^*_i(0)$, the hyperplane $H^*_i(\tau)\equiv \langle c_i(\tau),x^*(\tau)\rangle$ supports reachable set $\mathcal R^{\Phi}$.
\end{theorem}
\begin{proof}
The proof follows from Pontryagin's maximum principle \cite{hwang2005polytopic}. The contact point $x^*_i(\tau)$ evolves as $\dot{x^*_i} =  A(t) x(\tau) + B(t) u^*_i(\tau)$ for the given optimal control.
For the linear system $A(t),B(t)$, the costate $\lambda(t)$ evolves as $\dot{\lambda}(t)=A(t)^T \lambda(t)$.
Choose $c_i(\tau)=\lambda_i(\tau)$ where $\dot{\lambda_i}(\tau) = -A(\tau)^T \lambda_i(\tau)$ combined with the initial condition $\lambda_i(0)=\gamma^*_i(0)$ and suppress time indices for brevity.
Note the time derivative of $\langle \lambda_i, x \rangle$ equals:
\begin{equation*}
\begin{split}
&\frac{d}{d\tau}\langle \lambda_i, x \rangle = \langle \dot{\lambda_i}, x\rangle + \langle \lambda_i, \dot{x}\rangle \\
&= \langle -A^T \lambda_i, x\rangle + \langle \lambda_i, Ax + B u \rangle = \langle \lambda_i, u \rangle\leq \langle \lambda_i , u^*_i \rangle \\
\Rightarrow & \frac{d}{d\tau}\langle \lambda_i, x^*_i\rangle = \frac{d}{d\tau}\gamma^*
\end{split}    
\end{equation*}
Combined with the initial conditions on the points of contact, i.e., $\langle \lambda_i(0), x^*_i(0)\rangle = \gamma^*_i(0)$, one gets $\langle \lambda_i(\tau), x_i(\tau) \rangle\leq \langle \lambda_i(\tau) , x^*_i(\tau)\rangle = \gamma^*_i(\tau)$.
Hence, the hyperplane defined by $c^*_i$ and $x^*_i$ touches the reachable set.
\end{proof}

\begin{remark}
From \cite{varaiya2000reach}, the polytopic approximation can be made arbitrarily accurate. 
In fact, at time $\tau$:
\begin{equation*}
\text{convex hull}\{x^*_1,\cdots,x^*_{n_1}\} \subset \mathcal R^{\Phi}\blue{(\tau)} \subset \cap_{1}^{n_1}\{\lambda_i,x\}\leq\gamma^*_i
\end{equation*}
that is, the convex hull of the supporting points provide an under-approximation of the reachable set.
At the same time, the hyperplanes provide the over-approximation of the same.
\end{remark}

The polytopic reachable set approximation is a well-known numerically efficient method that is utilized by numerous reachable set computation tools.
The $\tau$-time reachable `tube' resulting from this method can be defined as the Minkowski sum $\oplus^{\tau}_{s=0} \reach{\Phi}{s}{\mathcal X_0,\Omega}$.
Combined with the DMDc-based method to obtain linear approximations of the NN model, this provides us with a scalable way to estimate reachable sets for NN models.
Clearly, the proposed numerical method relies on the universal approximation capabilities of NNs \cite{sonoda2017neural}.
Similarly, DMDc (more generally, DMD) converges in operator norm to the Koopman operator with increasing number of data points \cite{mezic2020numerical}.
Indeed, given enough data-points, and snapshot widths big enough, arbitrarily accurate reachable set approximations can be achieved.
This is true, in general, for most data-driven schemes.
In reality, the proposed framework provides a computationally cheap way to compute approximate reachable sets for the learned models, as the only additional computation steps involve matrix inversions in DMDc, and matrix exponentiation in propagating $\lambda$'s.

\begin{figure}[!t]
    \centering
    \includegraphics[width=1\columnwidth]{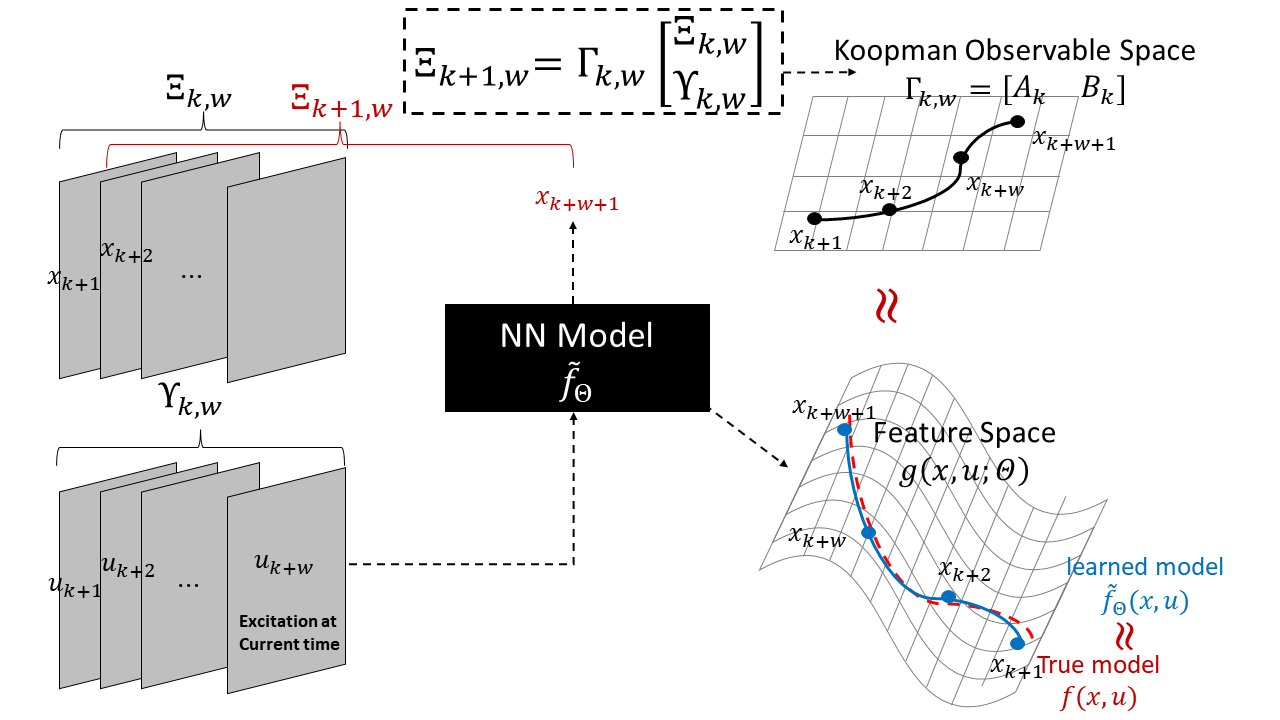}
    \caption{A comparison of the temporal data snapshots, as learned by the NN vs. the DMDc approximation}
    \label{fig:schematic-2}
\end{figure}

A depiction of the data snapshots, and the subsequent mappings discovered by the NN model and the DMDc method are shown in Fig.~\ref{fig:schematic-2}.
The learned model attempts to take temporal trajectory data to a higher-dimensional manifold, usually governed by the so-called ``feature space", parameterized by the NN parameters $\Theta$.
On the other hand, the DMD based method attempts to find approximations to the infinite dimensional Koopman operator, which considers the temporal trajectories in some ``observable space", where the trajectory evolution is (approximately) an action of a linear operator \cite{proctor2016dynamic, mauroy2020koopman}.
This is because DMD tries to find finite-dimensional truncations of the Koopman operator.
Hence, the proposed framework is amenable to a real-time implementation, and is presented in the subsequent section for a quadrotor.

\begin{remark}
Once the model is learned, the computational expense is of the order of $\mathcal O([\cdot]^{-1}) + \mathcal O (\exp{[\cdot]})\approx  \mathcal O (\exp{[\cdot]})$.
These can be accomplished in relatively computationally inexpensive ways by existing linear algebra libraries \red{(such as PyLops \cite{ravasi2020pylops} and Armadillo \cite{sanderson2010armadillo})}.

\end{remark}

\section{\blue{Reachable Sets for} a Quadrotor}
Although locally linear models of the quadrotor are often used for control synthesis \cite{bergamasco2014identification}, a fully  nonlinear quadrotor model is required to capture its wide dynamical range.
We demonstrate {our proposed} framework using a fully nonlinear, 12 degree-of-freedom {(DOF)} quadrotor model, based on \cite{wang2016dynamics}. 
In this section, we will look at the reachable set computation problem {of the 12DOF}, nonlinear dynamics, for a given $\mathcal X_0$ and $\Omega$ over time.

The state vector $\xi\in\reals^{12}$ is given by the 3D position $[x,y,z]^T$ and its respective velocities $[\dot{x},\dot{y},\dot{z}]^T$, and the 3D angular {attitude} $[\phi,\psi,\theta]^T$, and respective angular velocities $[\dot{\phi},\dot{\psi},\dot{\theta}]^T$.
The 12-DOF state $\xi$ evolves as follows:
\begin{align}\label{eq:quad-model-1}
\begin{bmatrix}
\ddot{x} \\ \ddot{y} \\ \ddot{z}
\end{bmatrix}
&= \begin{bmatrix}
-\frac{u_1}{m}\left(\sin{\phi}\cos{\psi} + \cos{\phi}\cos{\psi}\sin{\theta}\right) \\
-\frac{u_1}{m}\left(\cos{\phi}\sin{\psi}\sin{\theta} - \cos{\psi}\sin{\psi}\right) \\
g -\frac{u_1}{m}\left(\cos{\phi}\cos{\theta}\right)
\end{bmatrix}\\
\begin{bmatrix}
I_{xx}\ddot{\phi} \\ I_{yy}\ddot{\theta} \\ I_{zz} \ddot{\psi}
\end{bmatrix}
&= \begin{bmatrix}
u_2 \\ u_3 \\ u_4
\end{bmatrix} - \begin{bmatrix}
(I_{zz}-I_{yy})\dot{\theta}\dot{\psi}\\
(I_{xx}-I_{zz})\dot{\phi}\dot{\psi}\\
(I_{yy}-I_{xx})\dot{\theta}\dot{\phi}
\end{bmatrix}
\end{align}
where $I_{xx},I_{yy},I_{zz}$ are the moments of inertia along the 3 axes, $g$ is the acceleration due to gravity, and $m$ is the quadrotor's mass.
Variables $u_1,\cdots,u_4$ relate to the actual angular velocity command at the four rotors $\omega_1,\dots,\omega_4$ as:
\begin{equation}\label{eq:quad-model-2}
\begin{bmatrix}
u_1 \\ u_2 \\ u_3 \\ u_4
\end{bmatrix}
= \begin{bmatrix}
k_f & k_f & k_f & k_f \\
-lk_f & lk_f & lk_f & -lk_f \\
lk_f & lk_f & -lk_f & -lk_f \\
k_m & -k_m & k_m & -k_m
\end{bmatrix}
\begin{bmatrix}
\omega_1^2 \\ \omega^2_2 \\ \omega^2_3 \\ \omega^2_4
\end{bmatrix}    
\end{equation}
via the aerodynamic force and moment constants $k_f$ and $k_m$, respectively, and the distance of rotors from the center $l$ (see \cite{wang2016dynamics} for the derivation details).
Determining the force and moment constants $k_f$ and $k_m$, in itself, requires experimental system identification, let alone the nonlinear dynamics in (\ref{eq:quad-model-1}) through (\ref{eq:quad-model-2}).
As a result, learning the nonlinear dynamics for a quadrotor has been of interest in multiple works for control synthesis.

We assume that we have access to $n_T$ number of input-state trajectories $X_k,U_k$, each starting with a randomly initialized state $\xi_0 \sim \mathcal{U}_{\mathcal X_0}$.
We consider a NN capable of learning approximate dynamics $\tilde{f}_\Theta$ from the time series input-state trajectories $X_k,U_k$.
To this end, we employ a causal multistep NN to recover the nonlinear dynamics $\tilde{f}_\Theta$, based on \cite{raissi2018multistep}.
The key idea of a multistep NN is to use time series trajectory over a number of steps, say $\xi_k,\xi_{k-1},\dots,\xi_{k-m}$ and $\omega_{i,k},\cdots,\omega_{i,k-m}$ for $i=1,\cdots,4$, and find appropriate weights for a nonlinear function that takes the $m$-step trajectory and maps it to $\xi_{k+1}$ (see \cite{raissi2018multistep} for more details).
The actual functional approximation is offloaded to the multistep NN whose weights $\Theta$ minimize the mean-squared error over each $m$-step slice of the trajectory data, for a fixed $m$.

\red{Here, we employ a multistep neural network, which is a multi-layer perceptron NN based on \cite{raissi2018multistep}, that is used to perform system identification for differential equation based dynamical systems. 
The NN attempts to approximate the system dynamics $\dot{x}=f(x)$ from its trajectory data $\{x_k\}_{k=0}^N$ by `unrolling' a trajectory of length $N$ and approximating the dynamical map $\tilde{f}_\Theta:x_k \to x_{k+1}$ for $0\leq k\leq N-1$.
We used a multistep NN with 3 layers, including one hidden layer (12, 256 and 12 neurons, respectively).
The multistep scheme used was Adams-Moulton (i.e., using the trapezoidal rule to extend the function between $k$ and $k+1$) to train over 100 trajectories, with discretization time $\Delta T=0.1$s, activation function  $\tanh{(\cdot)}$, and mean square error (MSE) loss function with adaptive moment estimation (ADAM) as the optimizer.
The multistep NN took approximately 80 seconds to train, using Python, on an Intel Xeon CPU running at 2.20 GHz with a 13 GB memory and a 56 MB cache size, and was trained over 2,000 epochs and converges to an MSE loss of 1.26$\times 10^{-3}$.}

\begin{figure}[!t]
    \centering
    \includegraphics[width=0.6\columnwidth]{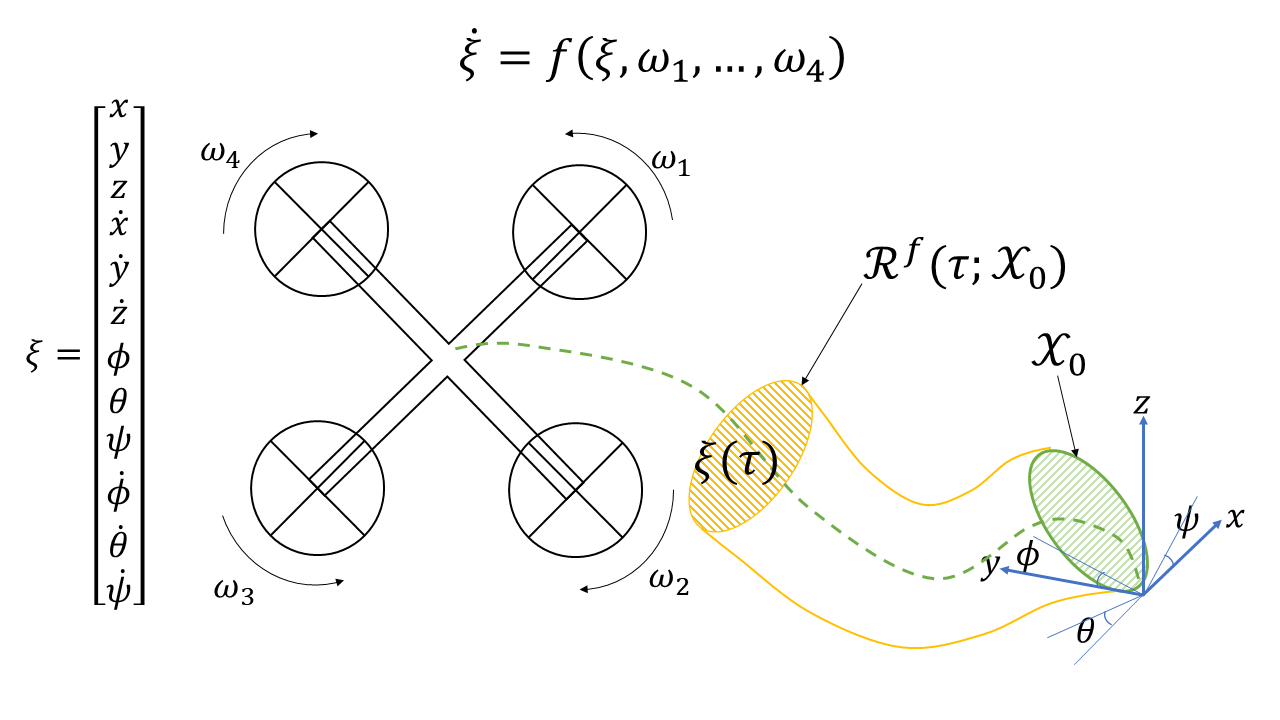}
    \caption{Reachable set computation problem for a quadrotor: state $\xi$ at time $\tau$ along some trajectory, starting from some initial set $\mathcal X_0$}
    \label{fig:quad-model}
\end{figure}

\subsection{Reachable Set Computation}
An example of the reachable set computation problem for the quadrotor model in (\ref{eq:quad-model-1})-(\ref{eq:quad-model-2}) is depicted in Fig.~\ref{fig:quad-model}.
A randomly chosen initial state starts from a given initial set (shown in green), and the collection of all possible evolutions of the quadrotor's state after time $\tau$ resides in the set $\reach{\tilde f}{\tau}{\mathcal X_0,\Omega,\Theta}$ (shown in yellow).
Additionally, the command input to each rotor has an additive noise as:
\begin{align}\label{eq:control-to-rotor}
\omega_i(t) &= v_i(t) + w_i, w_i(t)\sim\mathcal U_{[-0.25,0.25]} \text{ for }i=1,\cdots,4
\end{align}
and the admissible control set is defined as:
\begin{align}\label{eq:admissible-u}
\Omega &\triangleq\left\{\omega_1,\cdots,\omega_4\,\lvert\,\omega_i(t) = v_i(t) + w_i,\forall i \right\}
\end{align}
where the additive noise $w_i$ is assumed to be independent, identically distributed at all times $t$.
Equations (\ref{eq:control-to-rotor}) and (\ref{eq:admissible-u}) allow us to model the actuator noise into the reachable set approximation problem.
That is, $\reach{\tilde{f}}{\tau}{\mathcal X_0,\Omega,\Theta}$ contains all possible states that can be reached in time $\tau$, starting with $\xi_o\in\mathcal X_0$, under the noisy rotor command input set $\Omega$.

\begin{figure}[!htbp]
    \centering
    \includegraphics[width=0.65\columnwidth]{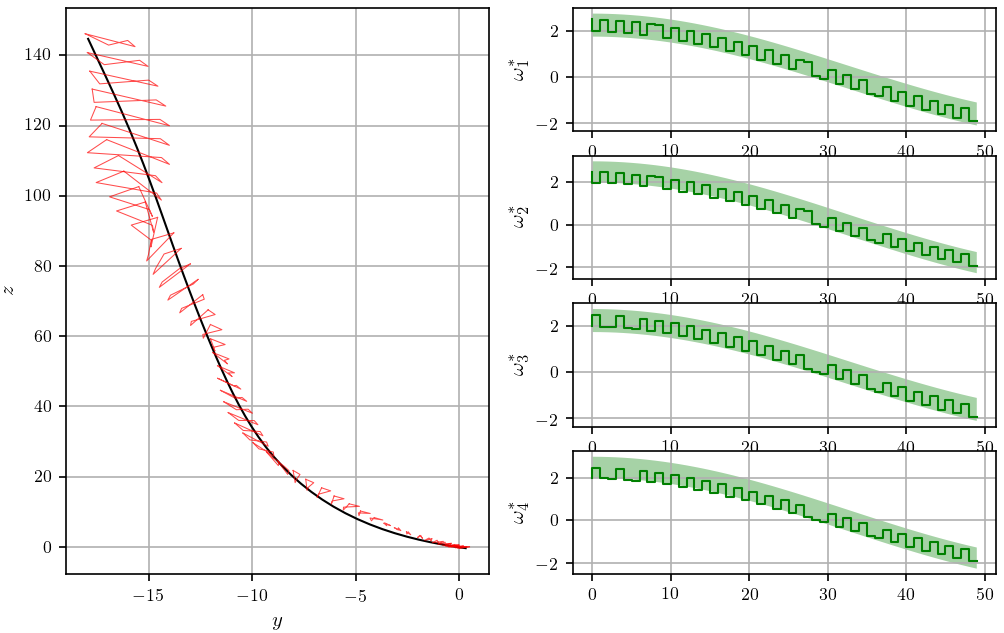}
    \caption{Inner approximations of Reachable Sets (convex hulls of points $\xi^*_i$) in the {$y-z$ plane}}
    \label{fig:yz-reach}
\end{figure}

We first learn the nonlinear model $\tilde{f}_\Theta$ for $n_T=100$ trajectories, each initialized with a random position $x,y,z\sim\mathcal{U}_{[-0.5,0.5]}$ coordinate (in m), and a random pose $\phi,\psi,\theta\sim\mathcal{U}_{[-0.1,0.1]}$ (in radians). 
This defines the initial set $\mathcal X_0$, and also provides explicit forms of hyperplanes $H_i$ at time $\tau=0$.
\red{To generate the training and testing datasets, we generated trajectories, sampled randomly to initiate from $\mathcal X_0$, and applied control sequences sampled randomly from $\Omega$ (as shown in Fig.~\ref{fig:yz-reach}) to each of the four rotors.}

\begin{figure}[!t]
    \centering
    \includegraphics[width=0.85\columnwidth]{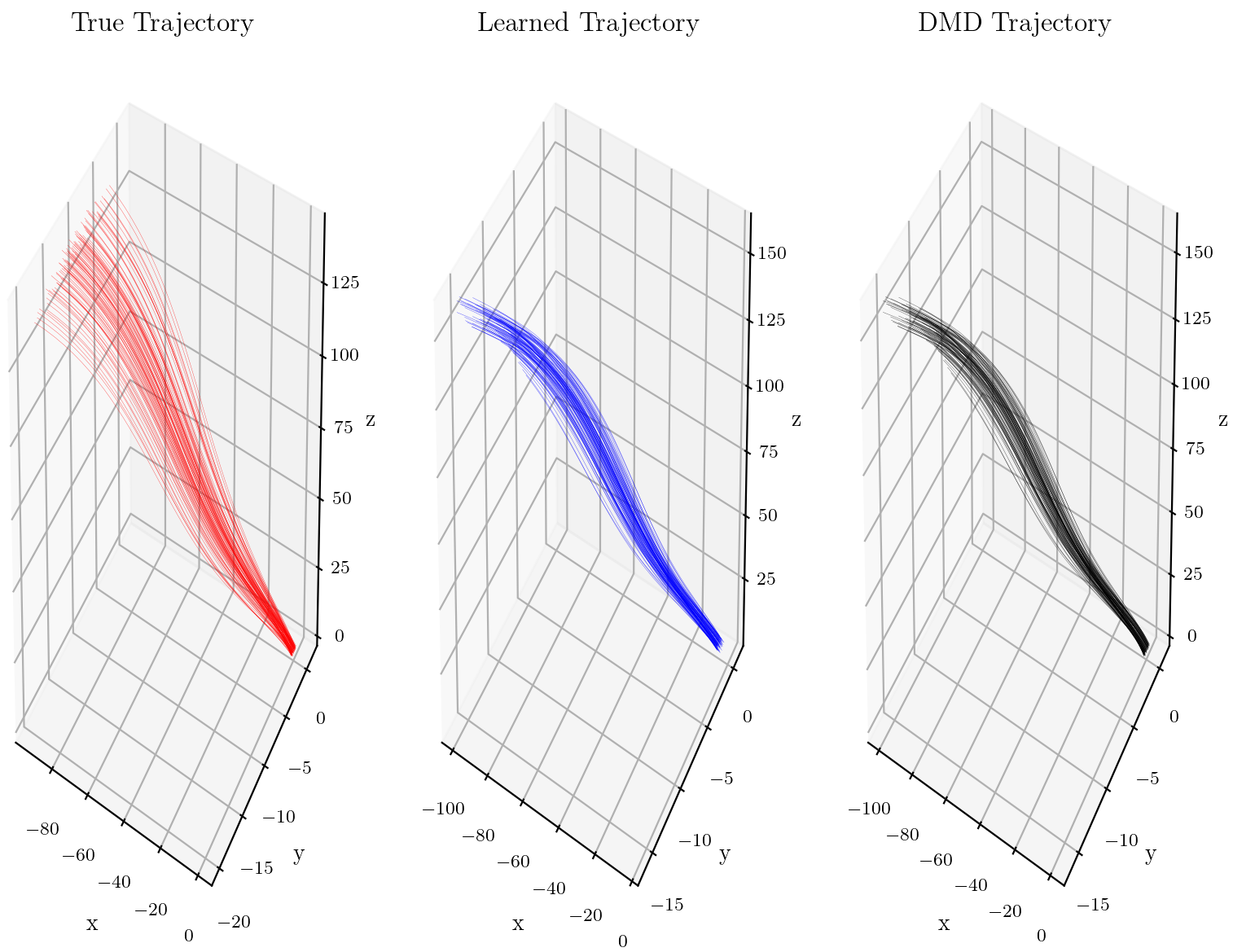}
    \caption{3D State trajectories: true trajectories (in red), trajectories reconstructed by the multistep NN (in blue), DMDc reconstructed trajectories (in black) }
    \label{fig:trajectories-nn-dmd}
\end{figure}

Based on the DMDc based framework in Section 3, {approximate linear time-varying models} $(A_k,B_k)$ are developed for $k=i\Delta t$ and a sampling time of $\Delta t=0.1$s and $k=0,\cdots,50$.
A comparison of the true, unknown position trajectories under mapping $f$, the predicted trajectories under learned model $\tilde{f}_\Theta$, and the trajectories reconstructed using DMDc under the mapping $A_k,B_k$ are shown in Fig.~\ref{fig:trajectories-nn-dmd}.
Clearly, given a relatively small data set $n_T=100$, the multistep NN is able to reconstruct the true trajectories accurately.
Additionally, despite the nonlinearities in the learned model $\tilde{f}_\Theta$, for the given small DMDc width $n_w= 8$, DMD reconstructions are also observed to be accurate.

\begin{figure}[!htbp]
    \centering
    \includegraphics[width=0.65\columnwidth]{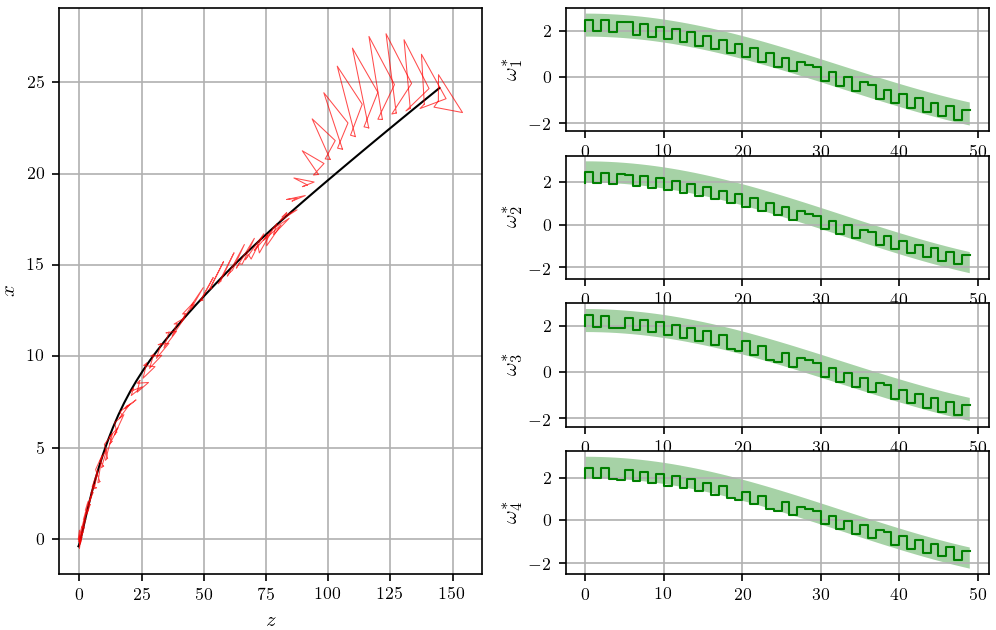}
    \caption{Inner approximations of Reachable Sets (convex hulls of points $\xi^*_i$) in the {$z-x$ plane}}
    \label{fig:zx-reach}
\end{figure}

Next, we represent the initial set $\mathcal X_0$ using $n_1=24$ hyperplanes and the admissible control set $\Omega$ using $n_2=8$ hyperplanes.
Each hyperplane supports $\mathcal X_0$ at a contact point $\xi^*_i$, and the $\tau$-time reachability problem becomes the $\tau$-time optimal control problem using the lifted system $A_k,B_k$.
That is, at a time $t=k\Delta t$, the lifted model $A_k,B_k$ is used to solve the optimal control problem in (\ref{eq:x_reach-contact}) and (\ref{eq:gamma_reach-contact}), propagating $\xi^*_i(k\Delta t)$ under the optimal control input $\omega_i^*(t)=\omega_i^*$ for $k\Delta t\leq t < k \Delta t + \tau$.
Propagating each contact point for time $t$ to $t+\tau$ gives the supporting structure for the $\tau$ time reachable set.
Given enough data, the multistep NN model approaches the true, unknown dynamical map, and the DMD lifted system approaches the learned model.
This accuracy in approximation is validated in Figs.~\ref{fig:yz-reach} and \ref{fig:zx-reach}.

The admissible control set $\Omega$ is given by $v_i(t)$ defined as a sinusoidal angular velocity inputs to the four rotors with an added noise, as shown in the green envelope in Figs.~\ref{fig:yz-reach} and \ref{fig:zx-reach}.
The solid green plots to the right in the figures depict the optimal control input that propagates an arbitrary hyperplane's contact point $\xi^*_i$ over time, by applying rotor control $\omega^*_1,\cdots,\omega^*_4$, as shown.
The points of contact of the hyperplanes are depicted in red, at each time {step} $0,\Delta t,2 \Delta t,...$ for a simulation duration of 5 seconds.
Inner approximations to the reachable tubes are simply Minkowski sums of the convex hulls in red.
An arbitrary trajectory starting from an $\xi(0)$ sampled from $\mathcal U_{\mathcal X_0}$ is reconstructed using the multistep NN, shown as a solid black line.

Despite the reasonable accuracy of the trained multistep NN (in Fig.~\ref{fig:trajectories-nn-dmd}), DMDc reconstruction finds accurate equivalent linear system models, and is therefore able to find approximate reachable sets that are fairly accurate, in real-time.
\red{
As noted in Section \ref{sec:intro-relatedworks}, there do not exist related methods that extend optimal control theory to obtain approximate reachable sets for NN models. 
The closest methods require exact, detailed NN models while employing LPs \cite{xiang2018reachability} and MILPs \cite{dutta2018output} to obtain reachable sets or output bounds.
As the proposed method does not require any internal details of NN architecture, treating it as a black box, a direct comparison is not very meaningful.
However, one can compare computational costs of the said methods as noted in the table below.
}

\begin{table}[ht]
\begin{center}
\red{
\caption{Comparing LP, MILP, and the proposed methods }}
\resizebox{0.7\columnwidth}{!}{
\begin{tabular}{l|c|c|c|c}
    \hline
    Method & NN & most expensive & \multicolumn{2}{c}{Computational Cost w.r.t.}\\
    \cline{4-5}
    & model & operation & \#layers & \#variables\\
    \hline
    LP method \cite{xiang2018reachability} & required & solve LP & $\mathcal O(L)$ & $\mathcal O(n^{\alpha})$ \\
    MILP method \cite{dutta2018output} & required & solve MILP & $\mathcal O(L)$ & exponential \\
    Proposed method & X & $\exp{[\cdot]}$ & $\mathcal O(1)$ & $\mathcal O(n^3)$ \\
    \hline
\end{tabular}
}
\end{center}
\label{table}
\end{table}
\red{Note that each layer $L$ in the NN architecture with $n$ variables introduces $\mathcal O(Ln)$ variables for the LP and MILP based methods. 
Since the MILP problems are NP-hard, the method in \cite{dutta2018output} has a worst case complexity as bad as brute force search, (hence an exponential worst case computational complexity). 
The LP-based method has a computational complexity of $\mathcal{O}[L(4nL)^{2.5}]$ (i.e., the computational cost of solving one LP for each layer).
On the other hand, the associated computational costs for the proposed method are based only on the cost of $\exp{[\cdot]}\sim\mathcal{O}[(n+n_h)^3]$ for $n_h$ hyperplanes.
Despite being computationally cheaper, the LP-based method provides hyperrectangular approximations to the reachable sets (hence, loose overapproximations), and is closer to interval reachable set methods such as \cite{thapliyal2022approximate,abate2022computing}.
Also, introducing the MILP and LP encodings requires extra overhead computations.
Finally, the propagation of the points of contact for each hyperplane can be done independently.
Therefore, the proposed method is more amenable to parallelization, as opposed to the inter-layer dependency of variables in the MILP and LP formulations.}

\subsection{Reachable set Computation under Rotor Failure}
Computing reachable sets under actuator failures is an important step to assess the compromised system's capabilities.
In this scenario, rotors 2 and 3 suffer from a total failure, and appear only as noise in the input channel.
This significantly modifies the admissible control set $\Omega$.
\begin{figure}[!htbp]
    \centering
    \includegraphics[width=0.85\columnwidth]{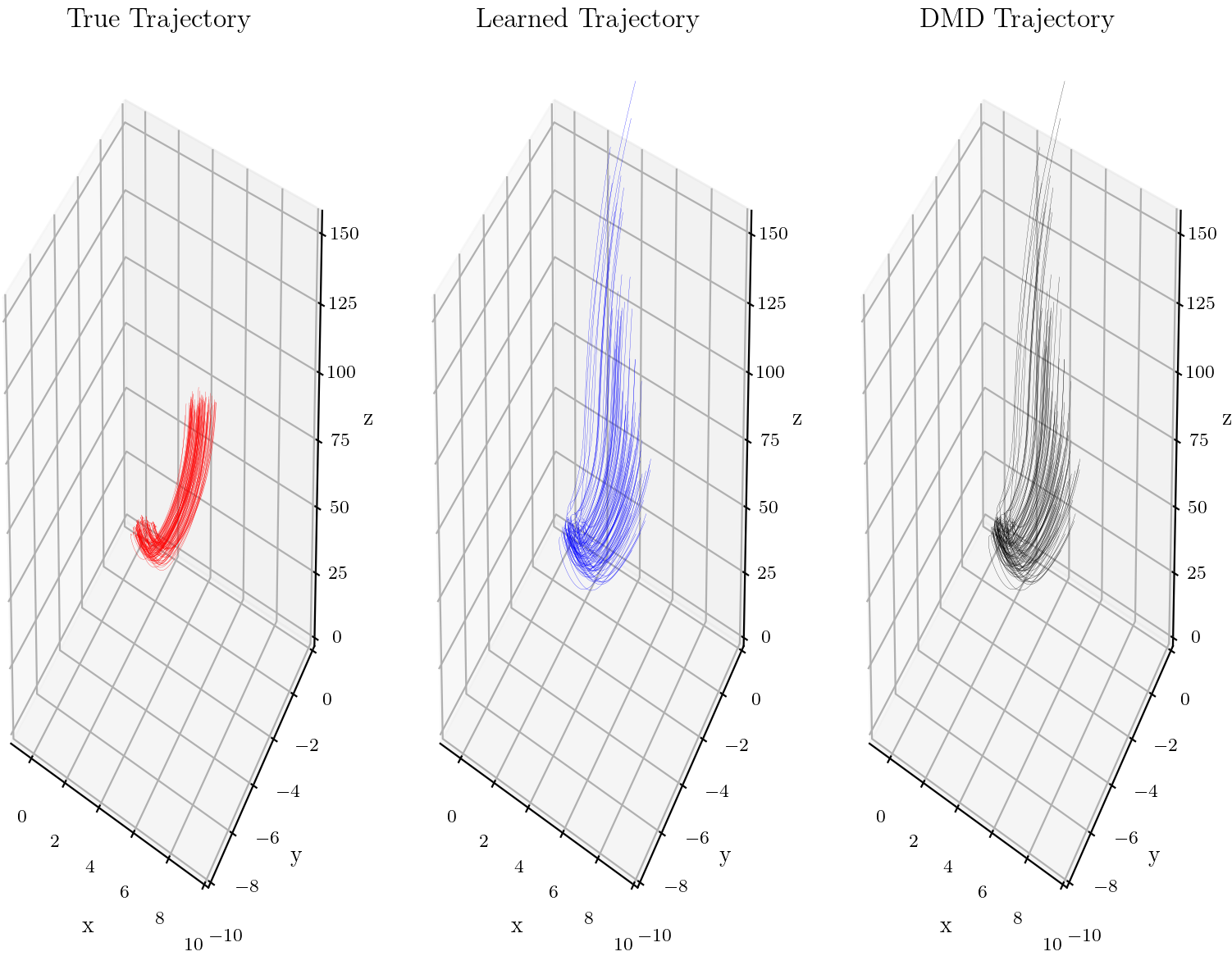}
    \caption{3D State trajectories under rotor failure at rotors 2 \& 3: true trajectories (in red), trajectories reconstructed by the multistep NN (in blue), DMDc reconstructed trajectories (in black)}
    \label{fig:trajectories-nn-dmd-fault}
\end{figure}
\begin{figure}[!htbp]
    \centering
    \includegraphics[width=0.65\columnwidth]{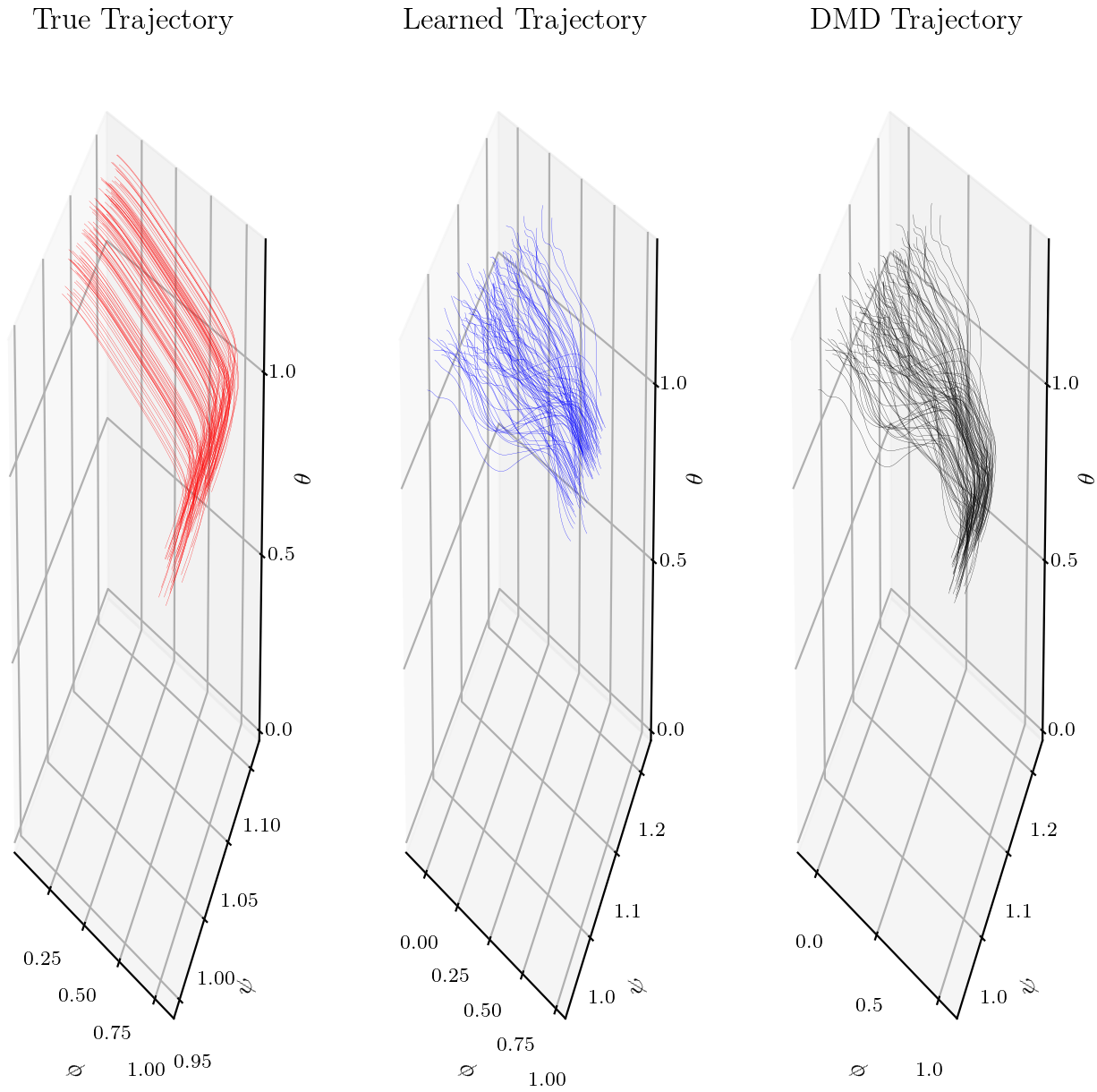}
    \caption{3D Attitude trajectories under rotor failure at rotors 2 \& 3: true trajectories (in red), trajectories reconstructed by the multistep NN (in blue), DMDc reconstructed trajectories (in black)}
    \label{fig:trajectories-nn-dmd-fault-theta}
\end{figure}
Due to the rotor failures, the NN model strays from the true trajectory, as shown in Fig.~\ref{fig:trajectories-nn-dmd-fault}.
This is further exaggerated in the $\phi,\psi,\theta$ trajectories, shown in Fig.~\ref{fig:trajectories-nn-dmd-fault-theta}.
Note that both the NN model and the DMDc model are not close to the true angular pose trajectories, but the DMDc model is still close to the learned model.

\begin{figure}[!htbp]
    \centering
    \includegraphics[width=0.85\columnwidth]{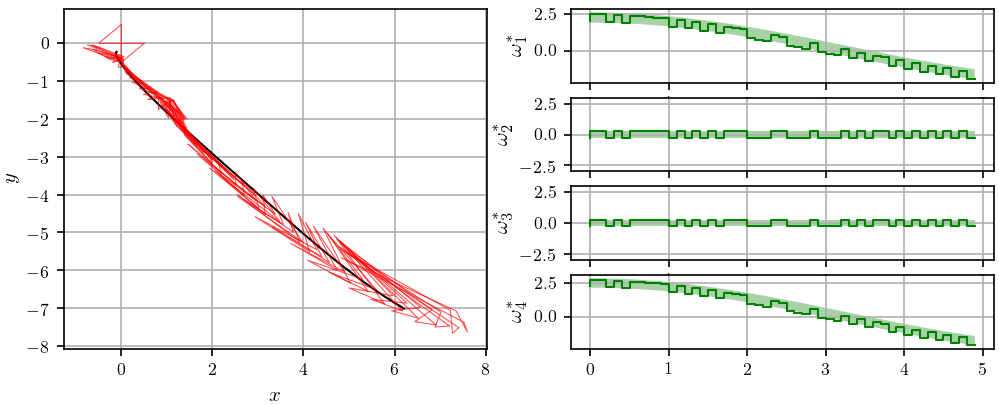}
    \caption{Rotor failure at rotors 2 \& 3: Inner approximations of Reachable Sets (convex hulls of points $\xi^*_i$) in the $x-y$ plane}
    \label{fig:xy-reach-failure}
\end{figure}
\begin{figure}[!ht]
    \centering
\includegraphics[width=0.85\columnwidth]{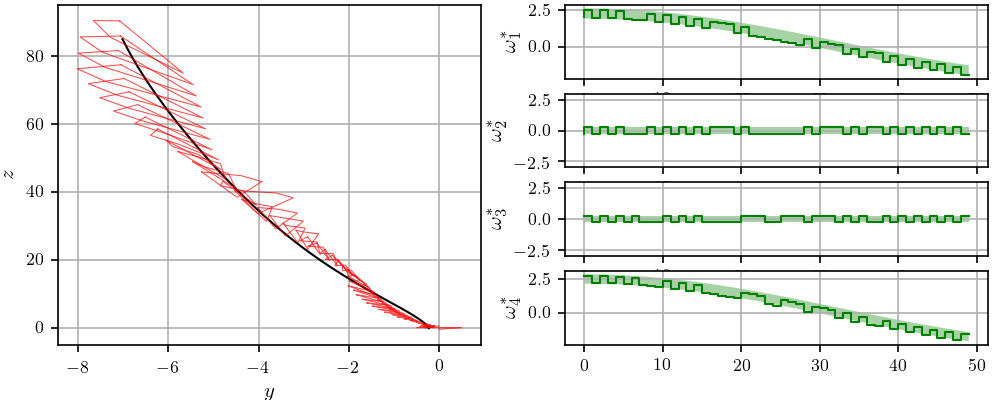}
\caption{Rotor failure at rotors 2 \& 3: Inner approximations of Reachable Sets (convex hulls of points $\xi^*_i$) in the {$y-z$ plane}}
    \label{fig:yz-reach-failure}
\end{figure}
\begin{figure}[!ht]
    \centering
    \includegraphics[width=0.85\columnwidth]{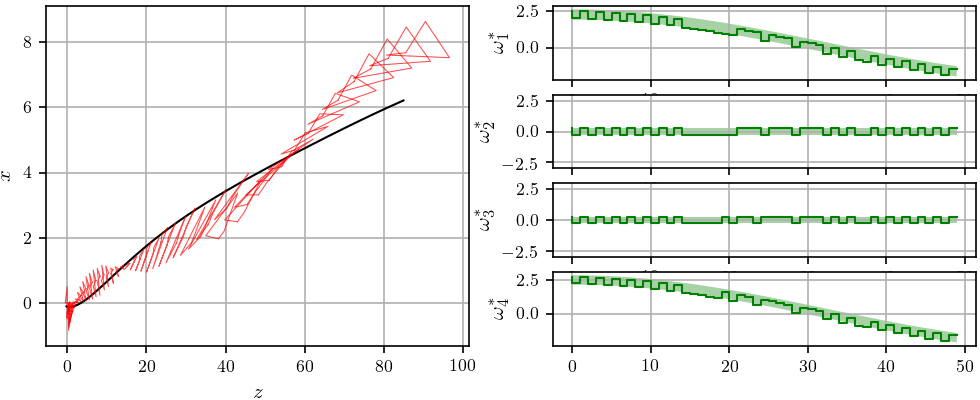}
    \caption{Rotor failure at rotors 2 \& 3: Inner approximations of Reachable Sets (convex hulls of points $\xi^*_i$) in the {$z-x$ plane}}
    \label{fig:zx-reach-failure}
\end{figure}
As a result of the rotor failure, the quadrotor's reachable sets are very different from the nominal case.
The admissible control sets $\Omega$ (shown as green envelopes in Figs.~\ref{fig:xy-reach-failure}, \ref{fig:yz-reach-failure}, and \ref{fig:zx-reach-failure}) are completely different due to $\omega_2$ and $\omega_3$ being bounded noise, while the remaining rotors provide the nominal angular velocity command.
This is also reflected in the resultant optimal control for an arbitrary hyperplane's point of contact, shown in the solid green lines.
The convex hulls of the points of contact denote inner approximations of the reachable sets.

Despite the highly nonlinear model, including rotor failures, and relatively small data set (both, $n_T$ and $n_w$), reachable set computation can be achieved with relatively high accuracy, and at a per-step computational increment of $<0.5$ seconds.
Therefore, the proposed method works well for real-time approximations of reachable sets.

\section{Conclusion}
In this paper\red{,} we presented a novel, data-driven framework for computing approximate reachable sets for nonlinear models learned using neural \red{networks.}
A computationally efficient lifting based method was proposed to find linear approximations of the learned model by exciting the neural network at each time step.
The proposed data-driven scheme was demonstrated to be computationally cheap, and was found to be amenable to usage in conjunction with other reachability tool sets.
Being data-driven, the proposed framework can be be made more accurate if more data are available. 
Moreover, real-time application of the framework was observed in a realistic quadrotor example.
Approximate reachable sets were computed for a causal neural network trained to learn the quadrotor's dynamics.
Additionally, modified reachable sets were computed for the quadrotor in a scenario that models rotor failures.
\blue{The proposed scheme was demonstrated to be of particular importance to safety critical scenarios where real-time approximation and update of reachable sets is required.}

As our immediate future work, we plan to investigate the space complexity of the framework and comment upon data required for a pre-described level of accuracy in reachable set computation. 
\red{Similarly, we plan to examine space complexity for a desired level of robustness against parameter variation.}
We also plan to look into developing efficient codes to facilitate a wrapper allowing plug-and-play usage with existing reachability toolboxes.

\section*{Acknowledgment}
{The authors would like to acknowledge that this work is partially supported by NSF CNS-1836952.}

\bibliographystyle{plain}  

\begin{thebibliography}{}
\bibitem{wang2016dynamics}Wang, P., Man, Z., Cao, Z., Zheng, J. \& Zhao, Y. Dynamics modelling and linear control of quadcopter. {\em 2016 International Conference On Advanced Mechatronic Systems (ICAMechS)}. pp. 498-503 (2016)
\bibitem{bergamasco2014identification}Bergamasco, M. \& Lovera, M. Identification of linear models for the dynamics of a hovering quadrotor. {\em IEEE Transactions On Control Systems Technology}. \textbf{22}, 1696-1707 (2014)
\bibitem{raissi2018multistep}Raissi, M., Perdikaris, P. \& Karniadakis, G. Multistep neural networks for data-driven discovery of nonlinear dynamical systems. {\em ArXiv Preprint ArXiv:1801.01236}. (2018)
\bibitem{proctor2016dynamic}Proctor, J., Brunton, S. \& Kutz, J. Dynamic mode decomposition with control. {\em SIAM Journal On Applied Dynamical Systems}. \textbf{15}, 142-161 (2016)
\bibitem{9072335}Alessandri, A., Bagnerini, P., Gaggero, M., Lengani, D. \& Simoni, D. Detection of Flow-Regime Transitions Using Dynamic Mode Decomposition and Moving Horizon Estimation. {\em IEEE Transactions On Control Systems Technology}. \textbf{29}, 1324-1331 (2021)
\bibitem{varaiya2000reach}Varaiya, P. Reach set computation using optimal control. {\em Verification Of Digital And Hybrid Systems}. pp. 323-331 (2000)
\bibitem{hwang2005polytopic}Hwang, I., Stipanović, D. \& Tomlin, C. Polytopic approximations of reachable sets applied to linear dynamic games and a class of nonlinear systems. {\em Advances In Control, Communication Networks, And Transportation Systems}. pp. 3-19 (2005)
\bibitem{sonoda2017neural}Sonoda, S. \& Murata, N. Neural network with unbounded activation functions is universal approximator. {\em Applied And Computational Harmonic Analysis}. \textbf{43}, 233-268 (2017)
\bibitem{mezic2020numerical}Mezic, I. On numerical approximations of the Koopman operator. {\em ArXiv Preprint ArXiv:2009.05883}. (2020)
\bibitem{devasia2017iterative}Devasia, S. Iterative machine learning for output tracking. {\em IEEE Transactions On Control Systems Technology}. \textbf{27}, 516-526 (2017)
\bibitem{thapliyal2021}Thapliyal, O. \& Hwang, I. Learning Based Cyberattack Design and Defense for Supervisory Control Systems. {\em 2021 European Control Conference (ECC)}. (2021)
\bibitem{romeres2019derivative}Romeres, D., Zorzi, M., Camoriano, R., Traversaro, S. \& Chiuso, A. Derivative-free online learning of inverse dynamics models. {\em IEEE Transactions On Control Systems Technology}. \textbf{28}, 816-830 (2019)
\bibitem{vaupel2020accelerating}Vaupel, Y., Hamacher, N., Caspari, A., Mhamdi, A., Kevrekidis, I. \& Mitsos, A. Accelerating nonlinear model predictive control through machine learning. {\em Journal Of Process Control}. \textbf{92} pp. 261-270 (2020)
\bibitem{cheng2019hidden}Cheng, J., Park, J., Cao, J. \& Qi, W. Hidden Markov model-based nonfragile state estimation of switched neural network with probabilistic quantized outputs. {\em IEEE Transactions On Cybernetics}. \textbf{50}, 1900-1909 (2019)
\bibitem{huang2021iterative}Huang, D., Yang, W., Huang, T., Qin, N., Chen, Y. \& Tan, Y. Iterative Learning Operation Control of High-Speed Trains With Adhesion Dynamics. {\em IEEE Transactions On Control Systems Technology}. (2021)
\bibitem{bansal2016learning}Bansal, S., Akametalu, A., Jiang, F., Laine, F. \& Tomlin, C. Learning quadrotor dynamics using neural network for flight control. {\em 2016 IEEE 55th Conference On Decision And Control (CDC)}. pp. 4653-4660 (2016)
\bibitem{ravasi2020pylops}Ravasi, M. \& Vasconcelos, I. PyLops—A linear-operator Python library for scalable algebra and optimization. {\em SoftwareX}. \textbf{11} pp. 100361 (2020)
\bibitem{sanderson2010armadillo}Sanderson, C. \& Others Armadillo: An open source C++ linear algebra library for fast prototyping and computationally intensive experiments. (Citeseer,2010)
\bibitem{chiuso2019system}Chiuso, A. \& Pillonetto, G. System identification: A machine learning perspective. {\em Annual Review Of Control, Robotics, And Autonomous Systems}. \textbf{2} pp. 281-304 (2019)
\bibitem{pillonetto2014kernel}Pillonetto, G., Dinuzzo, F., Chen, T., De Nicolao, G. \& Ljung, L. Kernel methods in system identification, machine learning and function estimation: A survey. {\em Automatica}. \textbf{50}, 657-682 (2014)
\bibitem{garg2017system}Garg, A., Tai, K. \& Panda, B. System identification: Survey on modeling methods and models. {\em Artificial Intelligence And Evolutionary Computations In Engineering Systems}. pp. 607-615 (2017)
\bibitem{lygeros2004reachability}Lygeros, J. On reachability and minimum cost optimal control. {\em Automatica}. \textbf{40}, 917-927 (2004)
\bibitem{ahn2020reachability}Ahn, H., Berntorp, K., Inani, P., Ram, A. \& Di Cairano, S. Reachability-based decision-making for autonomous driving: Theory and experiments. {\em IEEE Transactions On Control Systems Technology}. \textbf{29}, 1907-1921 (2020)
\bibitem{bansal2017hamilton}Bansal, S., Chen, M., Herbert, S. \& Tomlin, C. Hamilton-jacobi reachability: A brief overview and recent advances. {\em 2017 IEEE 56th Annual Conference On Decision And Control (CDC)}. pp. 2242-2253 (2017)
\bibitem{gayek1991survey}Gayek, J. A survey of techniques for approximating reachable and controllable sets. {\em [1991] Proceedings Of The 30th IEEE Conference On Decision And Control}. pp. 1724-1729 (1991)
\bibitem{dit2018reachability}Sandretto, J. \& Wan, J. Reachability analysis of nonlinear odes using polytopic based validated runge-kutta. {\em International Conference On Reachability Problems}. pp. 1-14 (2018)
\bibitem{1281781}Torrisi, F. \& Bemporad, A. HYSDEL-a tool for generating computational hybrid models for analysis and synthesis problems. {\em IEEE Transactions On Control Systems Technology}. \textbf{12}, 235-249 (2004)
\bibitem{althoff2018implementation}Althoff, M., Grebenyuk, D. \& Kochdumper, N. Implementation of Taylor models in CORA 2018. {\em Proc. Of The 5th International Workshop On Applied Verification For Continuous And Hybrid Systems}. (2018)
\bibitem{immler2018arch}Immler, F., Althoff, M., Chen, X., Fan, C., Frehse, G., Kochdumper, N., Li, Y., Mitra, S., Tomar, M. \& Zamani, M. ARCH-COMP18 category report: Continuous and hybrid systems with nonlinear dynamics. {\em Proc. Of The 5th International Workshop On Applied Verification For Continuous And Hybrid Systems}. (2018)
\bibitem{folkestad2021koopman}Folkestad, C. \& Burdick, J. Koopman NMPC: Koopman-based Learning and Nonlinear Model Predictive Control of Control-affine Systems. {\em ArXiv Preprint ArXiv:2105.08036}. (2021)
\bibitem{han2020deep}Han, Y., Hao, W. \& Vaidya, U. Deep learning of koopman representation for control. {\em 2020 59th IEEE Conference On Decision And Control (CDC)}. pp. 1890-1895 (2020)
\bibitem{dutta2018output}Dutta, S., Jha, S., Sankaranarayanan, S. \& Tiwari, A. Output range analysis for deep feedforward neural networks. {\em NASA Formal Methods Symposium}. pp. 121-138 (2018)
\bibitem{xiang2018reachability}Xiang, W. \& Johnson, T. Reachability analysis and safety verification for neural network control systems. {\em ArXiv Preprint ArXiv:1805.09944}. (2018)
\bibitem{huang2019reachnn}Huang, C., Fan, J., Li, W., Chen, X. \& Zhu, Q. Reachnn: Reachability analysis of neural-network controlled systems. {\em ACM Transactions On Embedded Computing Systems (TECS)}. \textbf{18}, 1-22 (2019)
\bibitem{xiang2018reachable}Xiang, W., Tran, H., Rosenfeld, J. \& Johnson, T. Reachable set estimation and safety verification for piecewise linear systems with neural network controllers. {\em 2018 Annual American Control Conference (ACC)}. pp. 1574-1579 (2018)
\bibitem{mauroy2020koopman}Mauroy, A., Mezić, I. \& Susuki, Y. The Koopman Operator in Systems and Control: Concepts, Methodologies, and Applications. (Springer Nature,2020)
\bibitem{thapliyal2022approximate}Thapliyal, O. \& Hwang, I. Approximate Reachability for Koopman Systems Using Mixed Monotonicity. {\em IEEE Access}. (2022)
\bibitem{abate2022computing}Abate, M. \& Coogan, S. Computing robustly forward invariant sets for mixed-monotone systems. {\em IEEE Transactions On Automatic Control}. (2022)


\end{thebibliography}

\end{document}